\documentclass[twocolumn]{aastex631}

\usepackage{bm}
\usepackage{url}
\usepackage{soul}
\newcommand\Tstrut{\rule{0pt}{2.6ex}}         

\begin{document}

\title{Astroparticles from X-ray Binary Coronae}

\author[0000-0002-5387-8138]{K.~Fang}
\affiliation{Department of Physics, Wisconsin IceCube Particle Astrophysics Center, University of Wisconsin, Madison, WI, 53706}

\author[0000-0001-6224-2417]{Francis Halzen}
\affiliation{Department of Physics, Wisconsin IceCube Particle Astrophysics Center, University of Wisconsin, Madison, WI, 53706}

\author[0000-0002-8433-8652]{Sebastian Heinz}
\affiliation{Department of Astronomy,  University of Wisconsin, Madison, WI, 53706}

\author[0000-0001-8608-0408]{John S. Gallagher}
\affiliation{Department of Astronomy,  University of Wisconsin, Madison, WI, 53706}
\affiliation{Department of Physics and Astronomy, Macalester College, St. Paul., MN,  55105}
 
\date{\today}

\begin{abstract}
The recent observation of high-energy neutrinos from the Galactic plane implies an abundant population of hadronic cosmic-ray sources in the Milky Way. 
We explore the role of the coronae of accreting stellar-mass black holes as such astroparticle emitters. We show that the particle acceleration and interaction timescales in the coronal region are tied to the compactness of the X-ray source. Thus, neutrino emission processes may similarly happen in the cores of active galactic nuclei and black hole X-ray binaries (XRB), despite of their drastically different masses and physical sizes. We apply the model to the well-measured XRB Cygnus~X-1 and find that the cascaded gamma rays accompanying the neutrino emission naturally explain the GeV emission that only presents during the source's hard state, while the state-averaged gamma-ray emission explains the LHAASO observation above 20~TeV. We show that XRB coronae could contribute significantly to the Galactic cosmic-ray and Galactic plane neutrino fluxes. Our model predicts variable high-energy neutrino emission from bright Galactic XRBs that may be observed by IceCube and future neutrino observatories. 
\end{abstract}


\section{introduction}

Neutrino emission from the Galactic plane was recently identified at the $4.5\sigma$ level of significance using ten years of data from the IceCube Observatory \citep{IceCubeGP}. The neutrino flux is consistent with diffuse neutrino emission by the cosmic-ray sea of the Milky Way but could also arise from a population of unresolved sources  \citep{Fang:2024nxn}. When converting the neutrino flux to a pionic gamma-ray flux, the IceCube Galactic plane emission is consistent with the Galactic diffuse emission (GDE) measured by {\it Fermi}-LAT at around 1~TeV~\citep{Fermi-LAT:2012edv} and the Tibet AS$\gamma$ Observatory above 100~TeV \citep{Tibet21}. Above $\sim 30$~TeV, it is comparable to, or higher than, the flux sum of the GDE and individual sources that are not associated with pulsars measured by the Large High Altitude Air Shower Observatory (LHAASO) \citep{Fang:2023ffx}. This implies two possibilities: 1) the GDE and a significant fraction of the LHAASO sources originate from hadronic interactions; or 2) there exists a population of $\gamma$-ray-obscured neutrino sources in the Milky Way.

Gamma-ray-obscured sources are evidently present in the extragalactic neutrino sky. The energy flux of the diffuse cosmic neutrino emission is higher than that of the isotropic $\gamma$-ray emission  \citep{2016PhRvL.116g1101M, Fang:2022trf}. This leads to a tension between the {\it Fermi}-LAT observation \citep{FermiIGRB} above $\sim 10$~GeV and the pionic $\gamma$ rays that leave the neutrino sources without attenuation and cascade down to GeV-TeV energies during their propagation in the extragalactic background light. Individual neutrino source observations also hint a source environment that is optically thick to GeV-TeV $\gamma$ rays. The neutrino flux of the active galaxy NGC~1068 is more than an order of magnitude higher than its GeV-TeV $\gamma$-ray flux \citep{IceCube:2022der}, suggesting that the neutrino emission site must be highly $\gamma$-ray-obscured.

The coronal region in the proximity of the central black hole provides an appealing site for both neutrino emission and $\gamma$-ray attenuation. Models of neutrino emission from the cores of active galactic nuclei (AGN) have long been proposed \citep{1981MNRAS.194....3B, PhysRevLett.66.2697, PhysRevLett.69.2738} and recently been developed to explain the NGC~1068 observations \citep{2020ApJ...891L..33I, PhysRevLett.125.011101, 2021Galax...9...36I, 2022ApJ...941L..17M}. 

Despite of the vast difference in scales, stellar-mass black hole binaries such as Cygnus~X-1 are close analogues to luminous AGNs as a result of their similar compactness \citep{2003MNRAS.345.1057M, 2004A&A...414..895F, 2015MNRAS.451.4375F}. The coronal compactness is defined as \citep{1983MNRAS.205..593G}:
\begin{equation}
    \ell = \frac{L_X}{R} \frac{\sigma_T}{m_ec^3} = 4\pi \frac{m_p}{m_e}
    \frac{R_g}{R}\frac{L_X}{L_E}
\end{equation}
where $L_X$ and $R$ are the luminosity and radius of the X-ray source, $R_g \equiv GM_{\rm BH}/c^2$ is the gravitational radius, $L_E = 4\pi GM_{\rm BH}m_p c / \sigma_T$ is the Eddington luminosity, $m_p$ and $m_e$ are the mass of proton and electron, respectively, and $\sigma_T$ is the Thomson cross section. For comparison, at $R = 10\,R_g$, the compactness of NGC~1068 is $\ell\approx 17$ with an intrinsic X-ray luminosity $4.6 \times 10^{43}\,\rm erg\,s^{-1}$ \citep{2015ApJ...812..116B} and a supermassive black hole mass of  $\sim 5\times 10^7\,M_\odot$ \citep{2015ApJ...802...98M}, very similar to that of the Galactic X-ray binary (XRB) Cygnus~X-1, which has $\ell \approx 19$ with an X-ray luminosity $2.2\times 10^{37}\,\rm erg\,s^{-1}$ during the hard state \citep{McConnell:2001yg} and a stellar mass black hole of $21.2\,M_\odot$ \citep{2021Sci...371.1046M}.

The shorter timescales of accretion on to black hole XRBs provide opportunities to study different accretion states and the associated high-energy radiation. Gamma rays have been observed from several XRBs, including LS~5039 \citep{HESS:2005wao, HESS:2006csb}, Cygnus~X-3 \citep{10.2307/27736637}), and Cygnus~X-1 \citep{2017MNRAS.471.3657Z, 4FGL}. Orbital modulation and absorption by the radiation of the companion star is observed, suggesting that the $\gamma$-ray emission must originate from a region smaller than the binary separation \citep{2013A&ARv..21...64D}. Neutrino emission from XRBs have previously been proposed in the context of jets interacting with matter of the wind or companion star \citep{1985Natur.316..418S, Romero:2003td, Bednarek:2005gf}. Below we focus on the canonical and well-measured Cygnus~X-1 and a scenario where its high-energy emission arises from the accretion flow.


\section{The Variable Cygnus~X-1}\label{sec:bg}
Cygnus~X-1 is a persistent black hole XRB at a distance of 2.22~kpc \citep{2021Sci...371.1046M}. Its X-ray emission varies on timescales of several weeks between two discrete levels: the ``hard state", during which the source spends $\sim 90\%$ of its time, characterized by a relatively low flux of soft X-rays ($\sim 2-10$~keV) and a relatively high flux of hard X-rays ($\sim 20-100$~keV); and the ``soft state", characterized by a relatively high soft X-ray flux and a relatively low hard X-ray flux (see Figure~\ref{fig:sed} and \ref{fig:bg} for the X-ray spectra during the two states).

The transition of the states is believed to relate to different configurations of the accreting matter \citep{McConnell:2001yg, Done:2007nc}. A general theoretical picture of the accretion flow structures in Cygnus~X-1 includes an inner hot, optically thin, geometrically thick accretion flow (the corona) and an outer cool, optically thick, geometrically thin disk.   
During the hard state, the source has low accretion rates. The accretion flow extends out to $\sim 100\,R_g$ and significantly contributes to the hard X-ray emission. Conversely, when the mass accretion rate increases, the flow becomes optically thick and collapses into a Shakura–Sunyaev disk. This explains why the blackbody radiation of the disk dominates the X-ray spectrum during the soft state.

While the soft and hard X-ray emission may be well reproduced by a multicolour disk model and hard thermal Comptonized spectrum, respectively, Cygnus~X-1 also presents a high-energy tail above the thermal-Compton spectrum starting at $\sim 0.1-1$~MeV. This component is explained as the scattering of nonthermal electrons accelerated by the accretion flow \citep{McConnell:2001yg, Romero:2014mna} or a compact jet \citep{Cangemi:2021dgt}. At even higher energies, $\gamma$-ray emission at $0.1-10$~GeV is detected from Cygnus~X-1 during its hard state \citep{2017MNRAS.471.3657Z, 4FGL}. Analyses of the {\it Fermi}-LAT data during its soft state only found upper limits above 100~MeV, but  obtained a tentative detection of a soft spectrum at the 40--80~MeV range \citep{2017MNRAS.471.3657Z}. 

Various models have been implemented to explain the $\gamma$-ray emission by Cygnus~X-1. All of them invoke relativistic electrons or protons produced in the source's steady or transient radio jets. In a leptonic scenario, electrons accelerated in the jets produce GeV emission by up-scattering the blackbody photons from the companion star \citep{2014MNRAS.442.3243Z}, the thermal-Compton emission by the accretion flow, or their own synchrotron radiation in the jets \citep{2013MNRAS.434.2380M, 2017MNRAS.471.3657Z}. These leptonic models are subjected to various constraints. For example, as the GeV flux of Cygnus~X-1 does not present a significant modulation \citep{2017MNRAS.471.3657Z}, the inverse Compton emission of stellar photons cannot dominantly contribute to the $\gamma$-ray emission. A synchrotron self-Compton model requires a highly clumped jet, while an external Compton model on disk photons needs a jet emission region very close to the accretion disk such that the synchrotron flux is comparable to the $\gamma$-ray flux \citep{2017MNRAS.471.3657Z}. 
Lepto-hadronic and hadronic scenarios have also been proposed, where non-thermal protons are accelerated in the jets in addition to electrons, producing GeV or TeV $\gamma$ rays by interacting with thermal protons or photons in the jet and in the stellar wind of the companion star \citep{Pepe:2015yxa, 2023A&A...673A.162P,Kantzas:2023}.

In this letter, we study an alternative scenario inspired by the modeling of extragalactic neutrino sources like NGC1068. We propose that relativistic protons are produced in the vicinity of the black hole, interact with the gas and photons in the coronal region, and produce neutrinos and $\gamma$ rays. In the hard state, 0.1-100~GeV $\gamma$ rays may partially survive as a result of the lower density of the disk photons. In the soft state, TeV $\gamma$ rays pair produce with the disk and coronal photons and cascade down to 1--100~MeV energies.  Our model naturally connects the variation of the MeV-GeV $\gamma$-ray flux to the transition of the accretion states. It also suggests that XRBs may be the hadronic accelerators that dominate the cosmic ray and high-energy neutrino emission of the Galaxy \citep{Sudoh:2022sdk,Sudoh:2023qrz}.

\section{Coronal Neutrino and $\gamma$-ray production}
\subsection{Proton Acceleration and Escape Timescales}
Proton acceleration in advection-dominated accretion outflows has long been suggested (for example see the review of \citealp{Yuan:2014gma}) and more recently studied in the context of NGC~1068 \citep{PhysRevLett.125.011101, Mbarek:2023yeq}.  Processes like turbulent dissipation may accelerate a fraction of electrons and protons in the two-temperature plasma of the corona into a non-thermal power-law distribution. In addition, a reconnection layer could form in the vicinity of a black hole, power the coronal X-ray emission by rapidly converting the magnetic energy into particle energy, and drive the hard/soft state transitions (e.g., \citealp{2014MNRAS.440.2185D}). Below we explore a simple scenario where particles get accelerated and interact inside the corona. The calculation would also be valid for a scenario where a population of protons get accelerated elsewhere and injected into the coronal region.

We model the corona as a spherical region with radius $R \equiv {\cal {R}}_{\rm rel}R_g$, where $R_g = 3.1\times 10^6\,(M_{\rm BH}/21.2\,M_\odot) \,{\rm cm}$, and assume that the coronal emission is subject to irradiation with soft photons from the outer accretion disk, which is assume to subtend a solid angle $\sim 2\pi$. The magnetic energy density in the  corona may be estimated as $u_B = B^2 / 8\pi = \xi_B\, u_X$, where $u_X = L_{d}/ 2\pi R^2c + L_c / 4\pi R^2c $ is the X-ray energy density radiated by the disk and corona. The scaling parameter $\xi_B \lesssim 1$ in regions dominated by magnetic turbulence \citep{Mbarek:2023yeq} and $\xi_B\sim 1/\beta_{\rm rec}$ in the case of magnetic reconnection, where $\beta_{\rm rec}$ is the plasma inflow speed in the reconnection layer in the unit of the speed of light, and $\beta_{\rm rec}\sim 0.1$ in the collionless relativistic regime \citep{Sironi:2015eoa}. The field strength in the corona of Cygnus~X-1 can then be estimated as $B \approx 1.2\times 10^6\, \xi_B ^{1/2} L_{X, 37.3}^{1/2} {\cal R}_{\rm rel, 1}^{-1} \,{\rm G}$.     
 
Comparisons of thermal Comptonization models with observations also suggest $B\lesssim 10^6$~G in the soft state and $B\sim 10^5-10^7$~G in the hard state \citep{2013MNRAS.430..209D}. 

\begin{figure}[t!]
    \centering
   \includegraphics[width=0.49\textwidth]{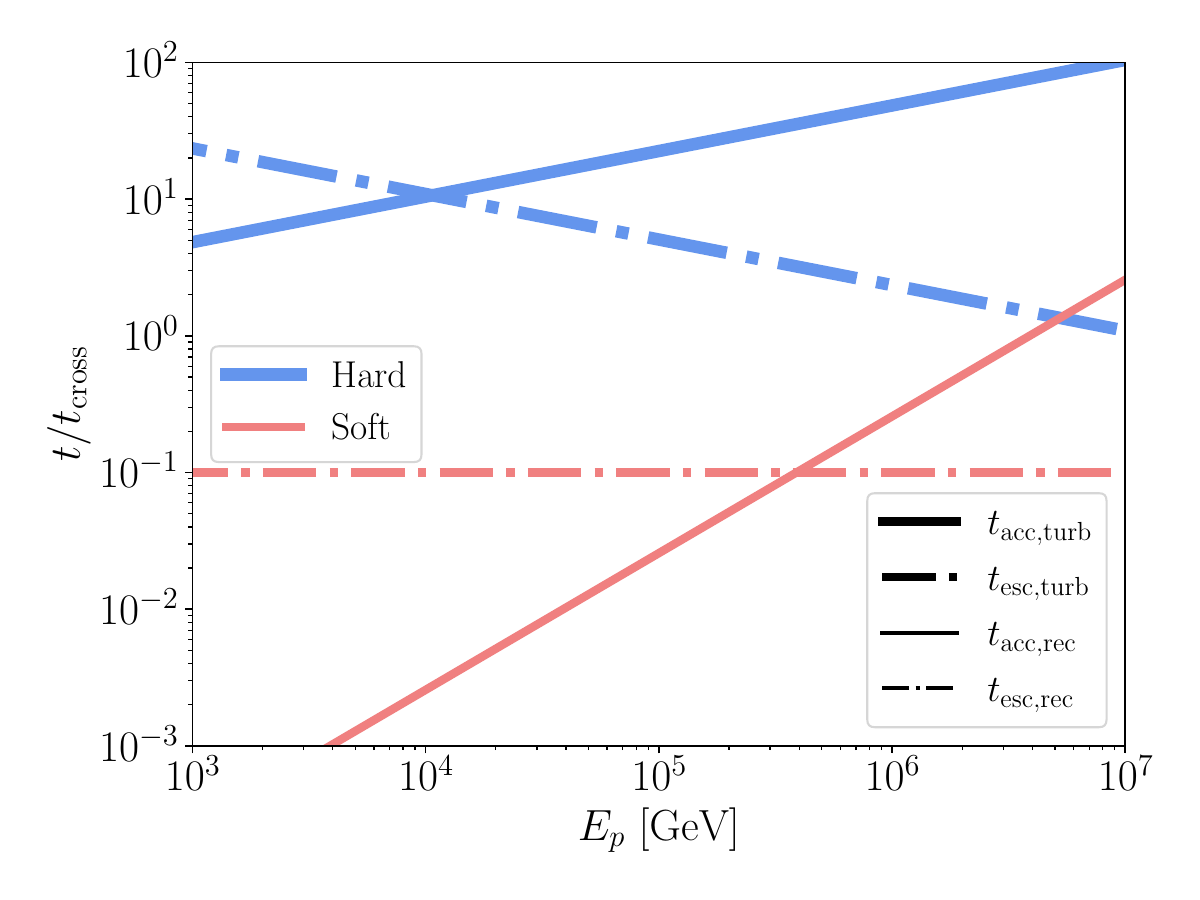}
    \caption{
    \label{fig:t_acc} Timescales of proton acceleration (solid curves) and escape (dash-dotted curves) in the units of light crossing time $t_{\rm cross}\equiv R/c$ as a function of the proton energy. In the hard state (blue), stochastic acceleration (equation~\ref{eqn:t_acc_turb} with $\eta = 10$) and particle diffusion (equation~\ref{eqn:t_esc_turb}) in a turbulent magnetic field are assumed to occur in a region of size $R = 100\, R_g$. The magnetic field strength is computed using the X-ray luminosity in Supplementary Table~1 for the assumed coronal size.  In the soft state (red), magnetic reconnection (equation~\ref{eqn:t_acc_rec} and \ref{eqn:t_esc_rec} with $\beta_{\rm rec} = 0.1$ is assumed. The ratios of the timescales to the light crossing time in this scenario do not depend on the coronal size.  } 
\end{figure}

The coronal electron number density is $n_e \approx \tau_T / \sigma_T R = 5\times 10^{16}\, \tau_T {\cal R}_{\rm rel, 1}^{-1}  \,\rm cm^{-3}$, where $\tau_T$ is the Thomson optical depth, and $\tau_T\sim 2$ in hard state and $\tau_T \sim 0.1$ in soft state \citep{McConnell:2001yg}. The pair magnetization may be parameterized as a function of the compactness,
\begin{eqnarray}
    \sigma_\pm = \frac{B^2}{4\pi n_e m_e c^2} =  \frac{\xi_B}{2\pi\,\tau_T}  \ell.
\end{eqnarray}
The proton magnetization is then $\sigma_p = {B^2}/{4\pi n_p m_p c^2} = \sigma_\pm n_e m_e / n_p m_p$. 
The proton number density is less well known. In a plasma where ions do not appreciably contribute to the mass density, $n_p m_p \ll n_e m_e$. In a mildly relativistic scenario $\langle\gamma_p\rangle \sim \sigma_p \sim 1$, $n_p m_p \sim n_e m_e$.

Stochastic proton acceleration may occur in the turbulent accretion outflow over a timescale $t_{\rm acc}^{\rm turb} \approx \eta (R/c)\, (c/v_A)^2\, (r_L/ R)^{2-q}$ \citep{PhysRevLett.125.011101}, where $\eta$ is a parameter related to the turbulence level $\delta B / B$, $r_L = E_p / eB$ is the Larmor radius,  $v_A =c (\sigma / (1+\sigma))^{1/2}$ is the Alfv\'en velocity, $\sigma = B^2 / (4\pi (n_p m_p c^2 + n_e m_e c^2))$ is the magnetization parameter, and $\sigma \sim \sigma_\pm$ in a plasma where ions do not dominate the mass density. Taking $q=5/3$ for Kolmogorov turbulence and comparing to a light crossing time $t_{\rm cross} \equiv R / c = 2.0\times 10^{-3}\, {\cal R}_{\rm rel, 1} \,\rm s$, we obtain

\begin{eqnarray}\label{eqn:t_acc_turb}
    \frac{t_{\rm acc}^{\rm turb}}{t_{\rm cross}} =\eta  \frac{1+\sigma}{\sigma}\left(\frac{E_p}{E_m}\right)^{1/3}
\end{eqnarray}
where $E_{m} \equiv eBR \approx 9.4\,B_6 {\cal R}_{\rm rel, 1} \,\rm PeV$. 
 
Accelerated particles diffuse away from the turbulent regions. The duration that protons are confined in the turbulent magnetic field can be estimated as $t_{\rm esc}^{\rm turb} \sim (R/c) (r_L / eBR)^{-1/3}$ \citep{2018ApJ...868L..28E, PhysRevLett.125.011101, Mbarek:2023yeq}, therefore,
\begin{equation}\label{eqn:t_esc_turb}
    \frac{t_{\rm esc}^{\rm turb}}{t_{\rm cross}} \approx  \left(\frac{E_p}{E_m}\right)^{-1/3}.
\end{equation}
In the absence of cooling processes, ions accelerated in the turbulent magnetic field may reach up to $t_{\rm acc, turb} < t_{\rm esc,turb}$, or $E_{p,\rm max}^{\rm turb} =  300\, B_6 {\cal R}_{\rm rel, 1}\,\eta_1^{-3/2}   (\sigma / 1+\sigma)^{3/2}\,\rm TeV$.

In the case that $\sigma \gg 1$, magnetic reconnection processes may plausibly occur \citep{Fiorillo:2023dts}. Protons may be accelerated in the magnetic reconnection layer over a timescale $t_{\rm acc}^{\rm rec}\approx E_p / \beta_{\rm rec}eBc$, or 
\begin{equation}\label{eqn:t_acc_rec}
   \frac{t_{\rm acc}^{\rm rec}}{t_{\rm cross}} = \frac{r_L}{R}\frac{1}{\beta_{\rm rec}}.
\end{equation}
Particles escape the magnetic reconnection region by advecting out of the reconnection layer after $t_{\rm ad} \approx L / c$, where $L\sim \beta_{\rm rec}R$ is the length of the reconnection layer \citep{Beloborodov:2017njh}. This yields
\begin{equation}\label{eqn:t_esc_rec}
    \frac{t_{\rm esc}^{\rm rec}}{t_{\rm cross}} = \beta_{\rm rec}.
\end{equation}
Comparing the acceleration and escape time leads to a maximum proton energy  $E_{p, \rm max}^{\rm rec} = E_m \beta_{\rm rec}^2 = 100\, B_6 {\cal R}_{\rm rel, 1}\,\beta_{\rm rec, -1}^2\,\rm TeV$.

In both turbulence and reconnection scenarios, protons may also escape the acceleration region by being advected to inside the innermost stable orbit. 
The fallback time is $t_{\rm fall} = R / \alpha v_K$, where $\alpha \sim 0.1$ is the viscous parameter and $v_K = (GM/R)^{1/2}$ is the disk velocity in circular Keplerian orbits  \citep{1973A&A....24..337S}, $t_{\rm fall}/t_{\rm cross} \approx 32  \, {\cal R}_{\rm rel, 1} ^{1/2}$. This timescale is usually longer, and thus less important, than the diffusion or advection time. A relativistic proton may be confined in the acceleration region by 
\begin{equation}\label{eqn:t_conf}
    t_{\rm conf} = \min(t_{\rm esc}, t_{\rm fall})
\end{equation}
with $t_{\rm esc}$ defined in equations~\ref{eqn:t_esc_turb} and \ref{eqn:t_esc_rec} for turbulent and reconnection scenarios, respectively. Figure~\ref{fig:t_acc} presents the acceleration and escape time for a Cygnus~X-1-like corona.


\begin{figure*} 
    \centering
   \includegraphics[width=0.9\textwidth]{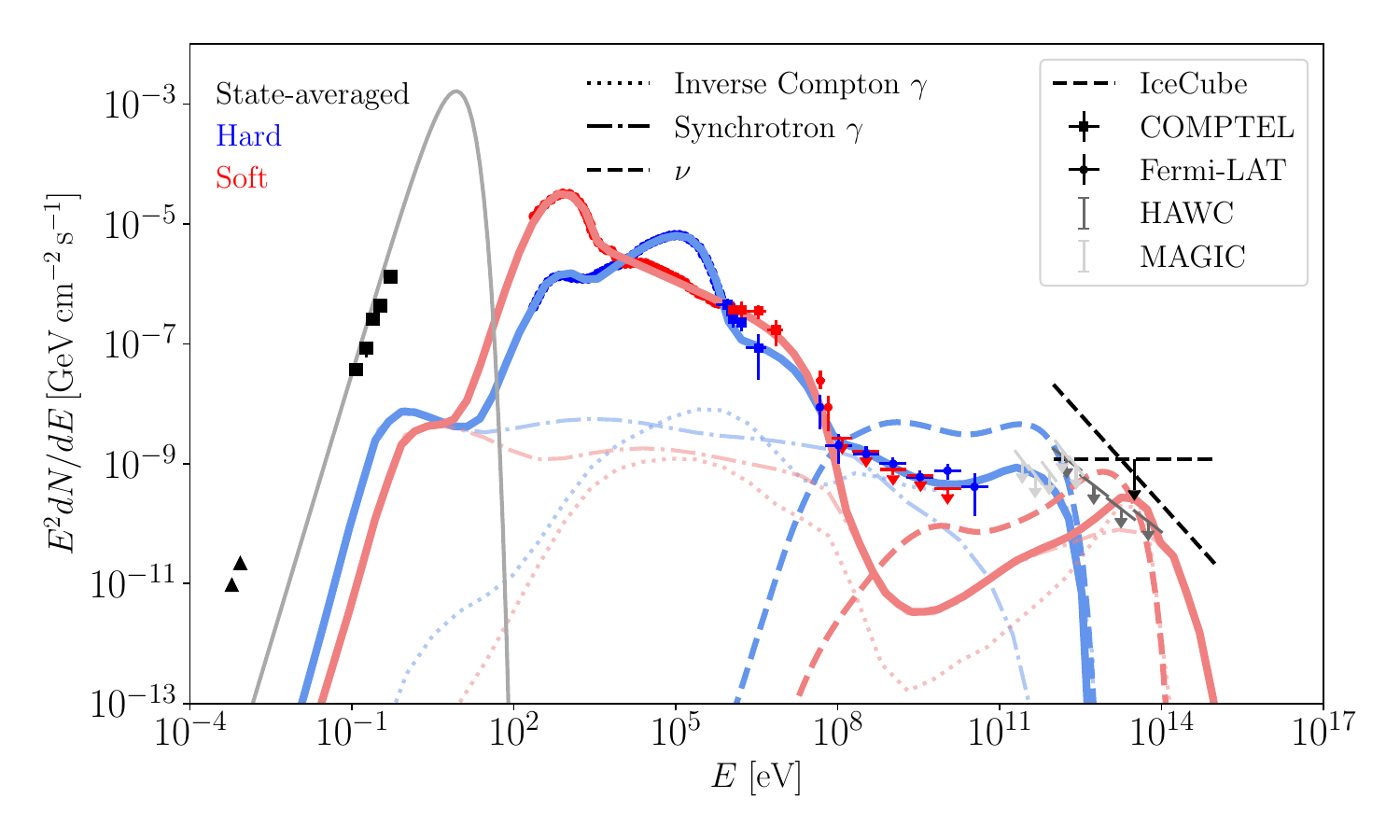}
    \caption{
    \label{fig:sed} Broadband spectral energy distribution of Cygnus~X-1 measured by BeppoSAX, INTEGRAL, COMPTEL, and {\it Fermi}-LAT \citep{DiSalvo:2000ji,Frontera:2000qi,Poutanen:2008gp,2017MNRAS.471.3657Z} during its hard (blue) and soft (red) states. Black and grey colors indicate state-averaged or state-insensitive observations in radio (triangle markers; \citealp{2001MNRAS.327.1273S, 2006MNRAS.369..603F}), infrared (square markers; \citealp{1980A&A....92..238P, 1996A&A...315L.113M}), gamma-ray by MAGIC (light grey limits; \citealp{MAGIC:2017clu}) and HAWC (grey limits; \citealp{HAWC:2021mdm}), and high-energy neutrinos by IceCube (black dashed limits corresponding to the 90\% C.L. median sensitivities for time-integrated searches using track-like events in ten-year IceCube data with both $dN/dE\propto E^{-2}$ and $E^{-3}$ assumptions and scaled to all-flavor flux; \citealp{IceCube:2019cia,IceCube:2022jpz}). The dotted, dash-dotted, and dashed curves correspond to the inverse-Compton radiation by secondary electrons and cascaded electrons, synchrotron radiation by electrons and protons, and neutrino emission by protons in the coronal emission models, respectively. The solid curves denote the total electromagnetic radiation in the two states by summing the background radiation (see Appendix~\ref{appendix:radbg}) and emission originated from nonthermal protons in the corona.    
}
\end{figure*}

\subsection{Interaction Timescales}
Relativistic protons cool by interacting with the ambient gas and radiation fields. The optical depths for protons due to photopion interaction ($p\gamma_b\rightarrow N\pi$) can be estimated as 
\begin{equation}\label{eqn:t_pg}
    \tau_{p\gamma}=\frac{t_{\rm conf}}{t_{\rm cross}}\frac{t_{\rm cross}}{t_{p\gamma}}\approx \frac{t_{\rm conf}}{t_{\rm cross}} \frac{\tau_T + 1}{\Omega} \frac{m_ec^2}{\epsilon_X}\frac{\sigma_{p\gamma}}{\sigma_T} \ell
\end{equation}
where $t_{p\gamma} \approx (n_X \sigma_{p\gamma} c)^{-1}$ is the $p\gamma$ interaction time, $\sigma_{p\gamma}$ is the inelastic photonpion interaction cross section, and the photon number density is estimated as $n_X \sim (\tau_T+1) L_X / \Omega R^2 c \epsilon_X$, where $\Omega \sim 4\pi$ if $L_X$ is dominated by the corona and $\Omega \sim 2\pi$ is $L_X$ is dominated by the disk. The same scaling relation applies to the Bethe-Heitler ($p\gamma_b\rightarrow pe^+e^-$) and pair production ($\gamma\gamma_b\rightarrow e^+e^-$) processes after replacing $\sigma_{p\gamma}$ by the inelastic Bethe-Heitler cross section $\sigma_{\rm BH}$ and pair production cross section $\sigma_{\gamma\gamma}$.

The proton-proton interaction ($pp\rightarrow N\pi$) time may be written as 
\begin{equation}\label{eqn:t_pp}
    \tau_{pp}=\frac{t_{\rm conf}}{t_{\rm cross}}\frac{t_{\rm cross}}{t_{pp}}=\frac{t_{\rm conf}}{t_{\rm cross}} \frac{n_p}{n_e}\frac{\sigma_{pp}}{\sigma_T}  \tau_T
\end{equation}
where $\sigma_{pp}$ is the inelastic proton-proton cross section. 

Equations~\ref{eqn:t_conf}, \ref{eqn:t_pg}, and \ref{eqn:t_pp} show that the optical depth for proton interaction in the corona region only depends on the dimensionless compactness parameters and the Thomson optical depth, thus the conditions for proton interaction and secondary production must be similar in the coronae of XRBs and AGNs, despite of their drastic differences in sizes and timescales. 

Figure~\ref{fig:tau} in Appendix~\ref{appendix:opticalDepth} presents the optical depth to relativistic protons and high-energy $\gamma$-rays computed using the radiation background of Cygnus~X-1.

\subsection{Gamma-ray and neutrino spectra}
We compute the neutrino and $\gamma$-ray spectra by iteratively solving a set of transport equations (as listed in Appendix~\ref{appendix:transport}) that account for the particle escape, energy loss and injection due to proton-proton, photopion, Bethe-Heitler, synchrotron, synchrotron self-absorption, inverse Compton, and photon-photon pair production processes.

For the hard state, we assume $R= 100\,R_g$, yielding $\sigma_\pm\sim 0.1$. Stochastic acceleration of protons from a thermal pool or a pre-accelerated population is possible in such a regime. We assume a proton injection spectrum that follows $Q_p^{\rm inj} = dN_p/dE_p\propto E_p^{-s}$ from the proton rest mass to $E_{p,\rm max}^{\rm turb}=10$~TeV. We adopt a spectral index $s = 2$ though the resulting secondary spectra are similar within the range $s\sim 2-2.3$. The diffusion of protons takes longer than the light crossing time and efficiently confines the particles for interactions. As a result, neutrinos may be produced in both $pp$ and $p\gamma$ processes. 

For the soft state, we assume $R = 30\,R_g$. The magnetization parameter $\sigma_\pm$ reaches $\sim 60$, making magnetic reconnection a favored scenario. The proton spectrum may follow a broken power-law with $dN_p/dE_p\propto E_p^{-1}$ below a break energy $E_{p,\rm br} \sim \sigma_p m_p c^2$ and $dN/dE_p \propto E_p^{-s}$ with $s>2$ \citep{Fiorillo:2023dts}. We thus inject a proton spectrum $Q_p^{\rm inj} \propto E_p^{-3}$ between  $E_{p,\rm min} = 60\,\rm GeV$ and $E_{p,\rm max}^{\rm rec}= 300$~TeV. Protons leave the reconnection layer rather fast and only $p\gamma$ interaction is efficient in this case.   

In both states we set the proton number density to $n_p m_p = n_e m_e$ and use the radiation background computed from the observed spectrum as described in Appendix~\ref{appendix:radbg}. The energy density of nonthermal protons is normalized to the X-ray energy density, $u_p = u_X$.

Figure~\ref{fig:sed} compares the resulting $\gamma$-ray and neutrino spectra to the broad-band spectral energy distribution of Cygnux~X-1. In the hard state, GeV and TeV $\gamma$ rays and electrons are produced by $p\gamma$ and $pp$ interactions. Most of them cascade down to 10-100~MeV energies, explaining the MeV $\gamma$-ray tail observed by COMPTEL. A fraction of the attenuated photons show up in the 1-100~GeV energy range as observed by LAT.  In the soft state, fewer interaction happens as a result of the poorer confinement of charged particles in the reconnection layer. Meanwhile, the high compactness makes the accretion flow opaque to $\gamma$ rays above $0.1$~GeV, and photons and pairs from $10-100$~TeV energies cascade down all the way to 10-100~MeV, which explains the observations of COMPTEL and LAT. The $\gamma$-ray spectrum, comparing with the observations by MAGIC, HAWC, and LHAASO, is presented in Figure~\ref{fig:gammaSED}.

Our model predicts a high-energy neutrino spectrum that cuts off at $\sim 1$~TeV during the hard state and peaks at $\sim 10$~TeV during the soft state.  The difference in these cutoff energies is mostly due to the difference in the maximum proton energies of the two states. In the hard state, neutrinos below and above $\sim 50$~GeV mostly come from $pp$ and $p\gamma$ processes, respectively, as suggested by the proton optical depth in Figure~\ref{fig:tau}.   In the soft state, neutrinos dominantly come from $p\gamma$ interactions. The neutrino spectra are consistent with the non-detection of IceCube for a point source at Cygnus~X-1's declination but could be observed in the future  with combined data samples from different event selections and across experiments.

\section{conclusions and discussion}\label{sec:dis}
Motivated by the $\gamma$-ray-opaque neutrino sources in the extragalactic sky, we propose that astroparticles may be produced in the accretion flow of Galactic black hole XRBs.  
We show that the acceleration and interaction timescales of relativistic protons are tied to the compactness parameter of an X-ray source. Therefore, the same cosmic-ray acceleration and neutrino production processes that happen in an AGN core may also occur in our local accreting black holes. 

We have assumed a spherical geometry for the coronal region. While the actual structure of a corona is highly unknown, its shape has been modeled as a slab, wedge, or cone in addition to a sphere. Recent polarimetric observations of Cygnus~X-1 start to reveal the geometric structure of the corona and the base of the jet \citep{Krawczynski:2022obw, 2024ApJ...960L...2C}. 
Moreover, we have assumed that the magnetic field strength, densities of the radiation field and the plasma are all constant within the coronal region. A more realistic picture would include a density gradient and a radially dependent magnetic field, though it would also introduce more free parameters to the model of particle acceleration and interaction. For example, we have verified that a similar solution like the one in Figure~\ref{fig:sed} may be found for a slab-like coronal region. We stress that the key results of this work, which only rely on the similarities in the compactness of AGN and XRB coronae, as well as the imprints of the accretion state transitions in the astroparticle production, remain robust under these uncertainties. 

Coronal protons may significantly contribute to the Galactic cosmic-ray population. 
The total X-ray luminosity of Galactic black hole XRBs is at the level of $L_X\sim (1-11)\times 10^{40}\,\rm erg\,s^{-1}$ \citep{2002A&A...391..923G, Heinz:2005hw, 2020MNRAS.493.3212C}. The coronae of these XRBs provide a total cosmic-ray power $L_{\rm CR, XRB} \lesssim L_X$.
This may explain the Galactic cosmic-ray spectrum, which requires $L_{\rm CR, obs} = u_{\rm CR} V_{\rm gal} / \tau_{\rm CR} \sim 5\times 10^{40}\,\rm erg\,s^{-1}$ when assuming cosmic-ray energy density  $u_{\rm CR}\approx 1\,\rm eV\,cm^{-3}$ above $1\,\rm GeV$, escape time $\tau_{\rm CR} \sim 3\,\rm Myr$ \citep{Blasi:2013rva} and volume of the Galactic disk $V_{\rm gal} \approx \pi R_{\rm gal}^2 d_{\rm gal}$ with $R_{\rm gal} \sim 15\,\rm kpc$ and $d_{\rm gal} \sim 0.2\,\rm kpc$. Unlike Cygnus~X-1 which spends most of its time in the hard state due to a high accretion rate, black hole XRBs are not commonly found to stay in the hard state. Considering some average of the soft and hard state, a cosmic-ray injection spectral index $s\sim 2.2-2.5$ could be expected, which  would explain the observed index $s_{\rm obs}\sim 2.7$ after the cosmic rays diffuse in the Galactic magnetic field. 

The optical depth for the relativistic protons is at the level of $\tau_{p\gamma} \sim 10^{-2}$ around $E_p\sim 100$~TeV, and depends on the size of the corona and confinement time at lower proton energy. Taking $\tau_{p\gamma}\sim 10^{-2}$, the total neutrino luminosity emitted by the XRB coronae is $L_{\nu, \rm XRB}\sim 3/8 L_{CR, \rm XRB} \tau_{p\gamma} \sim 6\times 10^{36}\,\rm erg\,s^{-1}$ above $1$~TeV. This corresponds to an all-flavor neutrino flux for an observer in the solar neighborhood  \citep{Fang:2024nxn} at the level of $E_\nu^2 F_\nu \approx 3 L_{\nu, \rm XRB} / 4\pi r_\odot^2= 1.3\times 10^{-6}\,\rm GeV\,cm^{-2}s^{-1}$ at 1~TeV, which is comparable to the  Galactic plane flux measured by IceCube \citep{IceCubeGP}.

\vspace{2em}

\begin{acknowledgments}
The work of K.F and F.H is supported by the Office of the Vice Chancellor for Research at the University of Wisconsin--Madison with funding from the Wisconsin Alumni Research Foundation. K.F. acknowledges support from the National Science Foundation (PHY-2110821, PHY-2238916) and NASA (NMH211ZDA001N-{\it Fermi}). This work was supported by a grant from the Simons Foundation (00001470, KF). K.F acknowledges the support of the Sloan Research Fellowship. The research of F.H was also supported in part by the U.S. National Science Foundation under grants~PHY-2209445 and OPP-2042807. J.S.G. thanks the University of Wisconsin College of Letters and Science for partial support of his IceCube-related research. SH acknowledges funding from NASA Astrophysics Theory Grant NNX17AJ98G.
\end{acknowledgments}

\bibliography{ref}


\appendix

\section{Radiation Background}\label{appendix:radbg}

We model the radiation field of Cygnus~X-1 from infrared to MeV as follows. 

Following \citet{McConnell:2001yg}, we model the emission by the corona and disk region as three components, including i) blackbody radiation from a Shakura–Sunyaev thin disk. We compute the emission using \texttt{agnpy} with a disk luminosity $L_{\rm d}$. The parameters of the inner and outer radii are selected to match the positions of the soft X-ray peak. ii) Inverse-Compton radiation by thermal electrons in the accretion flow, modeled as a power-law spectrum $dn_{\rm c, th}/d\epsilon\propto \epsilon^{-s_{\rm th}}\exp(-\epsilon / \epsilon_{\rm th, max})$ above a minimum energy $\epsilon_{\rm th, min}$. iii) Inverse-Compton radiation by non-thermal electrons in the accretion flow, modeled as a second power-law spectrum, $dn_{\rm c, nth}/d\epsilon\propto \epsilon^{-s_{\rm nth}}\exp(-\epsilon / \epsilon_{\rm nth, max})$ starting from the maximum energy of the thermal component $\epsilon_{\rm th, max}$. The third component is only included for the hard state. The geometry of the corona is modeled as a sphere, subject to soft photon flux from a source from the disk subtending a solid angle $\Omega\sim 2\pi$ and a source inside the corona subtending $\Omega \sim 4\pi$.

\begin{figure} 
    \centering
   \includegraphics[width=0.49\textwidth]{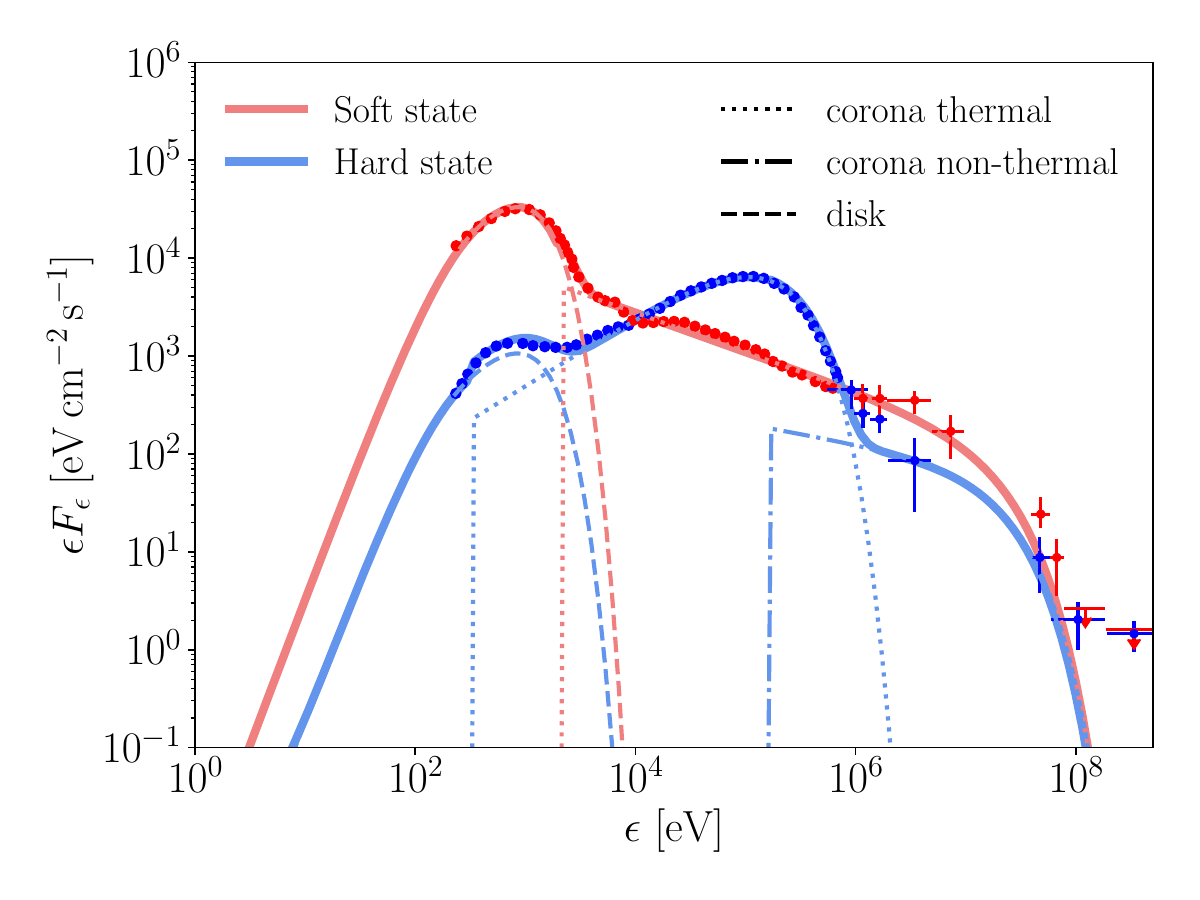}
    \caption{
    \label{fig:bg} Spectral energy distribution of Cygnus~X-1 from soft X-ray to MeV gamma-ray in the hard (blue) and soft (red) states \citep{DiSalvo:2000ji,Frontera:2000qi,Poutanen:2008gp,2017MNRAS.471.3657Z}.  The spectra at soft X-rays correspond to the unabsorbed, intrinsic emission. The dotted, dot-dashed, and dashed curves denote the emission from the thermal and non-thermal electrons in the accretion flow, and the accretion disk, respectively. The solid curves are the sum of all emission components.} 
\end{figure}

Figure~\ref{fig:bg} compares our model to the X-ray and MeV gamma-ray spectral energy distribution of Cygnus~X-1. We list the model parameters in Table~\ref{tab:table1}. They are in agreement with the parameters  obtained by fitting more sophisticated models to the X-ray data in the hard state and up to a factor of $\sim 2$ in the soft state \citep{McConnell:2001yg}. The difference is mostly due to the much simpler model we use. For example, our model does not account for the Compton reflection. Moreover, we model the hard X-ray and MeV gamma-ray emission as single power-laws, while \citet{McConnell:2001yg} computes the inverse Compton emission using ambient seed photons. 
Nonetheless, the details of the X-ray modeling barely matter here, as the high-energy neutrino and gamma-ray production is only sensitive to the energy density and spectral shape of the X-ray photons.

We model the stellar radiation as a blackbody spectrum with temperature $T_* = 2.55\times 10^4$~K \citep{2017MNRAS.471.3657Z} from a companion star with orbital separation $a_* = 3.2\times 10^{12}\,\rm cm$, with a radius $R_* = 19\,\rm R_\odot$ \citep{Pepe:2015yxa}. However, due to the large distance between the companion star and the black hole, the energy density of the stellar radiation at the corona is negligible compared to the emission by the disk and the synchrotron radiation by secondary electrons. 

\vspace{1em}

\section{Optical Depth}\label{appendix:opticalDepth}
\begin{figure}
    \centering
   \includegraphics[width=0.49\textwidth]{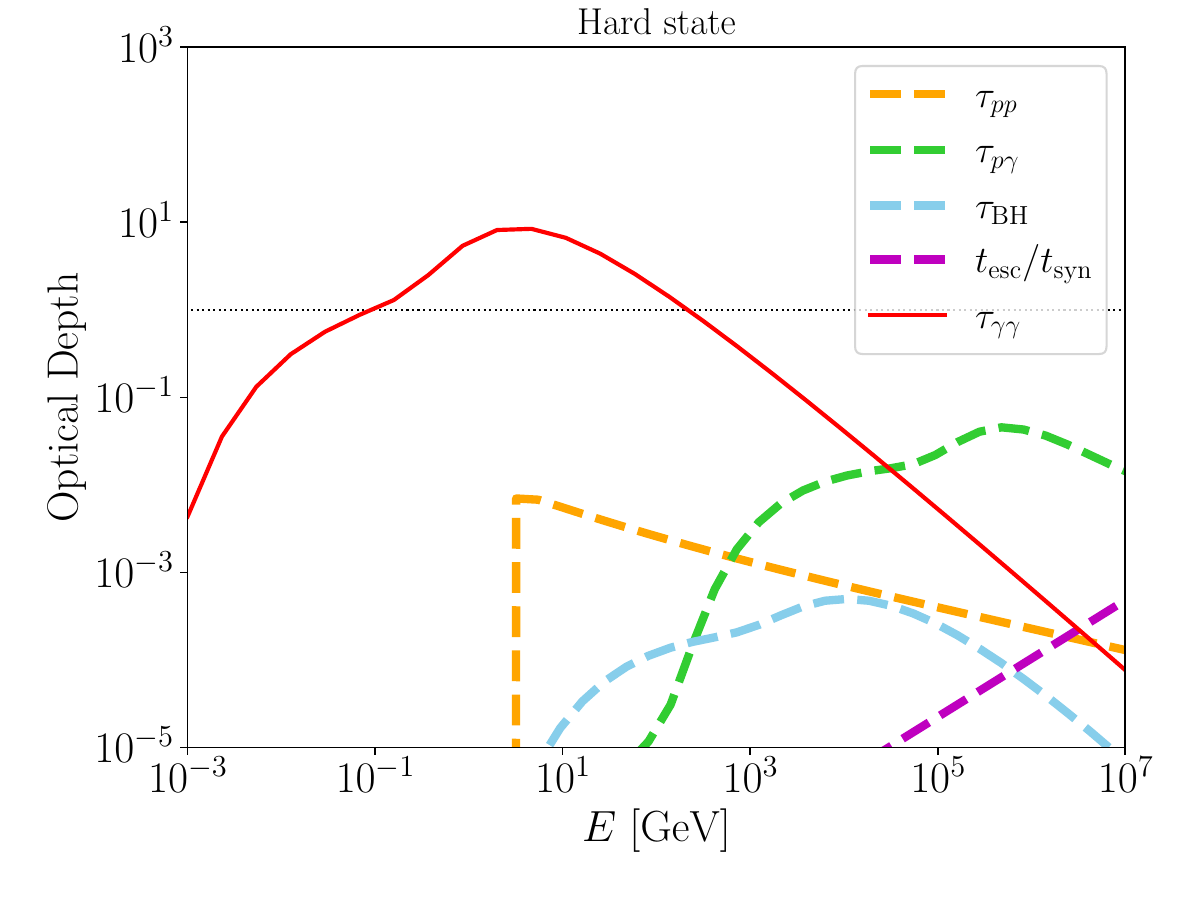}
   \includegraphics[width=0.49\textwidth]{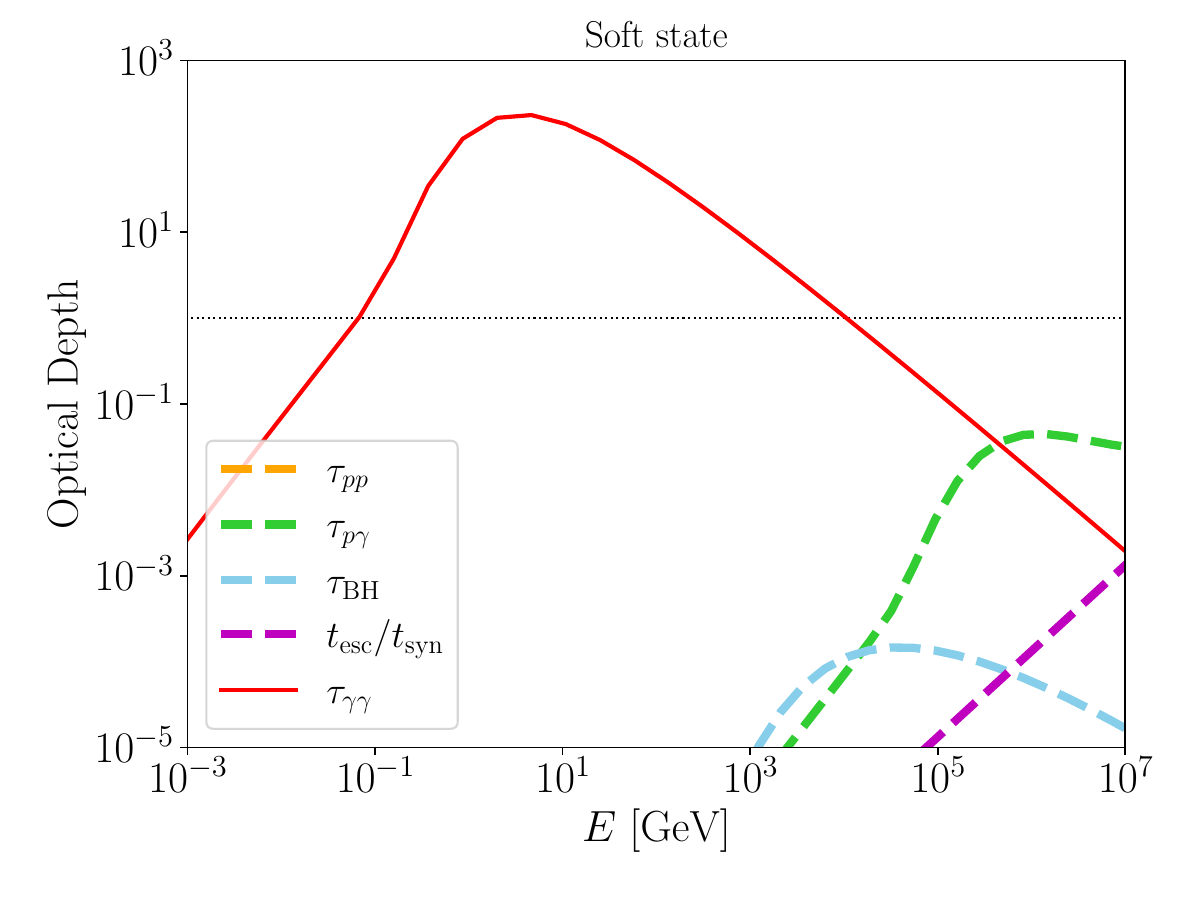}
    \caption{
    \label{fig:tau} Inelastic optical depth to protons (dashed curves) due to proton-proton interaction (orange), photopion production (green), Bethe-Heitler process (sky blue), and synchrotron radiation (magenta) as a function of the proton energy, and optical depth to $\gamma$-rays due to pair production (solid red curve) as a function of the $\gamma$-ray energy, during the hard (left) and soft (right) state of Cygnus~X-1. For reference, the black thin dotted line indicates an optical depth of 1. Parameters in use include $\xi_B = u_B / u_X = 1$,   ${\cal R}_{\rm rel}\equiv R / R_g = 100$ for hard state and 30 for soft state. For protons, the confinement time is assumed to be the diffusion time in the hard state and advection time in the soft state. The photon confinement time is the light crossing time $t_{\rm cross}$. The proton number density is set to $n_p m_p = n_e m_e$. The radiation background is the same as in Figure~\ref{fig:bg} which is computed using parameters in Table~\ref{tab:table1}. 
    }
\end{figure}

Using the radiation fields described in the last section, we compute the optical depth to relativistic protons and high-energy $\gamma$-rays due to various interaction processes. Figure~\ref{fig:tau} shows that in both states, the corona region is optically thick to $\gamma$-rays above $\sim10^{-1}$~GeV. The pionic $\gamma$-rays are attenuated by the infrared through  X-ray photons from the disk and cascade down to lower energies.

\section{Particle Transport Equations}\label{appendix:transport}
We compute the neutrino and $\gamma$-ray spectra by iteratively solving the following coupled transport equations. 
\begin{eqnarray}
  \frac{\partial N_p}{\partial t} + \frac{N_p}{\tau_{p,\rm esc}} + \frac{\partial}{\partial \gamma_p}\left[\left(\frac{d\gamma_p}{dt}\middle\vert_{\rm syn} + \frac{d\gamma_p}{dt}\middle\vert_{\rm BH} + + \frac{d\gamma_p}{dt}\middle\vert_{pp} + \frac{d\gamma_p}{dt}\middle\vert_{p\gamma}\right)N_p\right]  &=& Q_p^{\rm inj} \label{eqn:Np}  \\ 
    \frac{\partial N_e}{\partial t} + \frac{N_e}{\tau_{e,\rm esc}} + \frac{\partial}{\partial \gamma_e}\left[\left(\frac{d\gamma_e}{dt}\middle\vert_{\rm syn} + \frac{d\gamma_e}{dt}\middle\vert_{\rm IC}\right)N_e\right]  &=& Q_e^{p\gamma} + Q_e^{\rm BH} + Q_e^{pp} + Q_e^{\gamma\gamma} \\ 
    \frac{\partial N_\gamma}{\partial t} + \frac{N_\gamma}{\tau_{\gamma,\rm esc}} + N_\gamma c \left(a_{\gamma\gamma} + a_{\rm ssa}\right)   &=& Q_\gamma^{p\gamma}  + Q_\gamma^{pp} + Q_\gamma^{\rm IC} + Q_\gamma^{\rm e,syn}  + Q_\gamma^{\rm p,syn}\\ 
    \frac{\partial N_\nu}{\partial t} + \frac{N_\nu}{\tau_{\nu,\rm esc}}   &=&Q_\nu^{p\gamma}  + Q_\nu^{pp}  
\end{eqnarray}

Here $N_i \equiv (dN / dE)_i$ is the spectrum of a particle species  $i = p, e, \gamma, \nu$. 
For charged particles, the escape time $\tau_{\rm esc}$ is set to $t_{\rm conf}$ (equation~7) for a given proton or electron energy. For neutral particles,  $\tau_{\rm esc} = t_{\rm cross}$. The energy loss terms $d\gamma_i / dt\vert_{\rm process}$ and the injection terms $Q^{\rm process}_j$ denote the loss rate of particle species $i$ and the production of secondary particle $j$ due to one of the following process: synchrotron radiation (``syn"; \citealp{2012rjag.book.....B}), Bethe-Heitler process (``BH"; \citealp{1992ApJ...400..181C}), proton-proton interaction ($pp$; \citealp{PhysRevD.104.123027}), photonpion production ($p\gamma$; \citealp{Kelner:2008ke}), inverse Compton radiation (``IC"; \citealp{RevModPhys.42.237}), and pair production (also called as pair annihilation, $\gamma\gamma$; \citealp{Bottcher:1997kf}). For photons, the energy loss rates due to pair production and synchrotron self-absorption are computed using the efficiencies $a_{\gamma\gamma}$ \citep{1990MNRAS.245..453C} and $a_{\rm ssa}$ \citep{Dermer:2009her}. The equations are solved using a fully implicit finite difference scheme as in \cite{Stathopoulos:2023qoy} and evolved over a time period long enough to reach a steady-state solution.

The proton injection term $Q_p^{\rm inj}$ is assumed to be a single power-law from either the proton rest mass, in the turbulence scenario, or a break energy $\sim \sigma_p m_p c^2$, in the reconnection scenario, to the maximum energy. The actual nonthermal proton spectrum could be more complicated than a power-law for several reasons. We have neglected the momentum diffusion terms in Equation~\ref{eqn:Np} as the acceleration time is much shorter than the other timescales, though including such  terms is needed to capture the broadening of the particle distribution due to momentum space diffusion \citep{Becker:2006nz, Stawarz:2008sp}. Moreover, the acceleration process could be more complicated than what has been assumed.  For example, \citet{Mbarek:2023yeq} proposes an alternative acceleration picture to the turbulence and reconnection scenarios. They suggest that protons are injected into the corona with a hard spectrum and get re-accelerated to 1-10~TeV energies. Such a two-step acceleration is needed to explain the neutrino flux of  NGC~1068, as neither relativistic magnetic reconnection nor turbulence seems to accelerate sufficient relativistic protons if their initial energy follows a thermal distribution.

\section{Gamma-ray spectrum}\label{appendix:gammaSED}
\begin{figure} 
    \centering
   \includegraphics[width=0.49\textwidth]{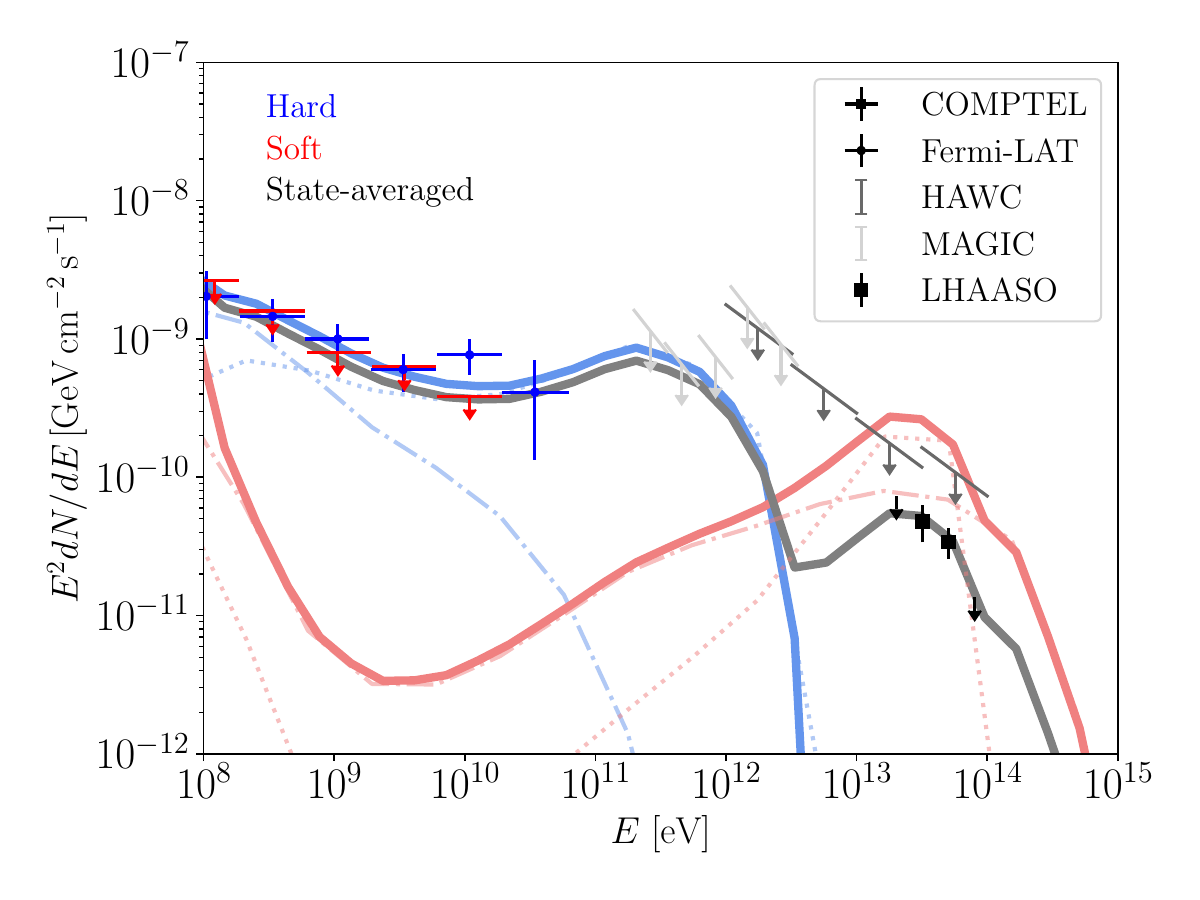}
    \caption{
    \label{fig:gammaSED} Same as Figure~\ref{fig:sed} but focusing on the gamma-ray spectrum of Cygnus X-1. The black curve corresponds to a state-averaged emission by assuming that source spends 80\% and 20\% of time in hard and soft states, respectively. The black square markers indicate a time-integrated observation by LHAASO \citep{Collaboration:2024wte}. 
    }
\end{figure}

Figure~\ref{fig:gammaSED} zooms into the spectral energy distribution of Cygnus X-1 in the gamma-ray energy range and compares our model to observations. The black curve indicates a state-averaged spectrum considering that on average the source spends $\sim 80\%$ and $\sim 20\%$ in hard and soft states, respectively \citep{Grinberg:2013qpa}. 
Our model explains the LHAASO observation above 20~TeV. We note that the emission between $\sim$10~GeV and $\sim$10~TeV would be further absorbed by the stellar photons \citep{2021ApJ...912L...4A}, resulting in periodic modulation.

\begin{table}
\caption{Parameters of the X-ray emission model, assuming a black hole mass of $21.2\,M_\odot$ and source distance $2.22\,\rm kpc$.   
\label{tab:table1} 
}
\begin{ruledtabular}
\begin{tabular}{ccccccccc}
  State  &  $L_{\rm c, th}$   & $s_{\rm th}$  & $\epsilon_{\rm th,  min}$  &  $\epsilon_{\rm th, max}$  &   $L_{\rm c, nth}$ &  $s_{\rm nth}$ & $\epsilon_{\rm nth, max}$ & $L_{\rm d}$  \\ 
  & $[\rm erg\,s^{-1}]$ & &  [keV] & [keV] & $[\rm erg\,s^{-1}]$    &    &    [keV] &  $[\rm erg\,s^{-1}]$ \\
  \hline  
 Hard    &   $2\times10^{37}$ & 1.3   &   $0.3$ &   $150$ &   $5\times 10^{35}$    &   $2.2$ & $2\times 10^4$	& $4\times 10^{36}$   \Tstrut\\
Soft   &   $1.2\times10^{37}$ & 2.4  &   $2$ &   $2\times10^4$ &   -    &  - & -	& $1.2\times 10^{38}$  \\
\end{tabular}
\end{ruledtabular}
\end{table}

\end{document}